\documentclass[twocolumn,prd,showpacs]
              {revtex4}

\usepackage{graphicx}

\begin{document}

\title{Black hole remnants due to GUP or quantum gravity?}

\author{Michael~Maziashvili}

\email{maziashvili@hepi.edu.ge} \affiliation{Department of
Physics, Tbilisi State University, 3 Chavchavadze Ave., Tbilisi
0128, Georgia \\Institute of High Energy Physics and
Informatization, 9 University Str., Tbilisi 0186, Georgia }

\begin{abstract}
Based on the micro-black hole \emph{gedanken} experiment as well
as on general considerations of quantum mechanics and gravity the
generalized uncertainty principle (GUP) is analyzed by using the
running Newton constant. The result is used to decide between the
GUP and quantum gravitational effects as a possible mechanism
leading to the black hole remnants of about Planck mass.
\end{abstract}

\pacs{03.65.-w, 04.60.-m, 04.70.-s, 04.70.Dy}


\maketitle

The cogent argument for the black hole to evaporate entirely is
that there are no evident symmetry or quantum number preventing
it. Nevertheless, the heuristic derivation of the Hawking
temperature with the use of GUP prevents a black hole from
complete evaporation, just like the prevention of hydrogen atom
from collapse by the uncertainty principle \cite{ACS}. The
generalized uncertainty relation takes into account the
gravitational interaction of the photon and the particle being
observed. This consideration relies on classical gravitational
theory \cite{AS,MS}. The quantum corrected Schwarzschild
space-time obtained with the use of running Newton constant also
indicates that the black hole evaporation stops when its mass
approaches the critical value of the order of Planck mass
\cite{BR}. On the other hand the quantum corrected gravity
modifies this GUP as well. It is fair to ask whether the halt of
black hole radiation is provided by GUP or it is due to quantum
gravitational effects. Let us give a critical view of this
problem.

Let us briefly discuss the modification of Heisenberg uncertainty
principle due to gravitational interaction. The main conceptual
point concerning GUP is that there is an additional uncertainty in
quantum measurement due to gravitational interaction. We focus on
consideration of this problem presented in \cite{AS}, ($\hbar=c=1$
is assumed in what follows). The approach proposed in this paper
relying on classical gravity is to calculate the displacement of
electron caused by the gravitational interaction with the photon
and add it to the position uncertainty. The photon due to
gravitational interaction imparts to electron the acceleration
given by $a=G_0\Delta E/r^2$ ($G_0$ is experimentally observed
value of Newton's constant for macroscopic values of distances).
Assuming $r_0$ is the size of the interaction region the variation
of the velocity of the electron is given by $\Delta v \sim
G_0\Delta E/r_0$ and correspondingly $\Delta x_g \sim G_0\Delta
E$. Therefore the total uncertainty in the position is given by
\begin{equation}\label{genunp}\Delta x\geq {1\over 2\Delta
E}+\alpha G_0\Delta E~,\end{equation} where $\alpha$ is the factor
of order unity in respect with the stringy induced GUP \cite{Ve}.
The Eq.(\ref{genunp}) exhibits the minimal observable distance.
However, the minimal observable distance is determined rather due
to collapse of $\Delta E$ than merely by the Eq.(\ref{genunp}),
because it puts simply the bound on the measurement procedure. In
this way one gets that for $\alpha \geq 2$ the $\Delta
x_{min}=\sqrt{2\alpha G_0}$ while for $\alpha < 2$ the minimal
observable distance is given by $\Delta
x_{min}=\sqrt{8G_0/(4-\alpha)}$. In the framework of this
discussion one can obtain the GUP in higher dimensional case as
well as on the brane \cite{MMN}.

The heuristic derivation of black hole evaporation proceeds as
follows \cite{ACS}. (In paper \cite{ACS} $\alpha=1$ is assumed).
The black hole is modelled as an object with linear size equal to
the two times the gravitational radius $2r_g$ and the minimum
uncertainty condition is assumed for the radiation. Then the lower
value of $\Delta E$
\begin{equation}\label{gupht}\Delta E={r_g - \left(r_g^2-G_0\right)^{1/2}\over 2G_0}~,\end{equation} that comes from
uncertainty relation is identified to the characteristic
temperature of the black hole emission with the constant of
proportionality $1/2\pi$. As one sees from Eq.(\ref{gupht}) the
GUP amended Hawking temperature becomes complex if the mass of the
black hole is less than $1/2G_0^{1/2}$, leading thereby to the
nonzero minimal black hole mass \cite{ACS}.

As it is evident the GUP assumes two $\Delta E$ values for a given
$\Delta x$. But the choice of the lower value is well motivated
physically because for relatively small values of $\Delta E$ the
gravitational uncertainty becomes negligible in comparison with
the standard term and therefore in the limit $\Delta x \gg
\sqrt{G_0}$ one has to recover the standard uncertainty relation.
On the other hand such a choice is motivated by the correct
asymptotic dependance of Hawking temperature on the black hole
mass in the framework of this heuristic approach.

However for the black hole with a radius not too far above the
Planck length the quantum fluctuations of the metric play an
important role. The effective Newton constant obtained by means of
the Wilson-type effective action has the form \cite{BR}

\begin{equation}\label{runnewcon}G(r)={G_0r^3\over
r^3+2.504G_0\left(r+4.5G_0M\right)}~,\end{equation} where $M$ is
the mass of the source. The effective Newton constant
Eq.(\ref{runnewcon}) depends on the mass of source and
correspondingly the mass of the test particle is implied to be
negligibly small in comparison with the source mass. Since the
gravitational uncertainty becomes appreciable when the energy of
photon approaches the Planck scale, which in turn exceeds very
much the mass of the standard model particles, one can safely use
the Eq.(\ref{runnewcon}). An interesting observation made in
\cite{BR} is that for $M$ less than the critical value
$M_{cr}=3.503\,G_0^{-1/2}$ the horizon disappears. So that the
black hole evaporation process comes to a complete halt when the
mass reaches to the critical value. By taking the quantum
corrected equation for gravitational radius as \cite{BR}
\[r_g=2G(r_g)\Delta E~,\] one finds the following expression for outer
horizon
\begin{widetext}
\begin{eqnarray}\label{ograd}r_g/G_0\Delta E&=&0.667 +
 0.265  \left(16 - 139.5\,t +10.392\sqrt{t(t-0.204 )(176.392 +
 t)}\right)^{1/3} \nonumber\\&-& {0.41(3t-4)\over \left(16 - 139.5 \,t +
10.392\sqrt{t(t-0.204)(176.392+t)}\right)^{1/3}}~,\end{eqnarray}\end{widetext}
where $t\equiv 2.504m_p^2/\Delta E^2$ and $m_p\equiv G_0^{-1/2}$.
The critical value of mass below which the horizon disappears is
given by $t_{cr}=0.204,~~~\Delta E_{cr}=3.503G_0^{-1/2}$.
Following the paper \cite{MS} one can combine this gravitational
radius and standard position uncertainty as
\begin{equation}\label{rncgenunp}\Delta x\geq \left\{
\begin{array}{ll}1/2\Delta E ~~~~~~~~~ \mbox{if}~~~~ \Delta E\leq 3.503G_0^{-1/2}
\\r_g\left(\Delta E\right)~~~~~~~\mbox{if}~~~~ \Delta E > 3.503G_0^{-1/2}\end{array}\right.~,\end{equation}
where $r_g(\Delta E)$ is given by Eq.(\ref{ograd}). From
Eq.(\ref{rncgenunp}) one gets the following minimal length $\Delta
x_{min}=0.143\,G_0^{1/2}$. The linear combination
\begin{equation}\label{qucgenunp}\Delta x\geq {1\over 2\Delta
E}+r_g(\Delta E)~,\end{equation} does not make any sense for
$\Delta E< \Delta E_{cr}$ since for these values of energy there
is no horizon at all. (Let us comment that the idea of paper
\cite{MS} to determine the total uncertainty for $\Delta E>\Delta
E_{cr}$ by Eq.(\ref{qucgenunp}) is not quite clear because the
collapse of $\Delta E$ puts simply the limitation on the
measurement). One has to define the gravitational disturbance of
the electron position directly. Taking the quantum corrected
potential around the photon to be $-\Delta E G(r)/r$ then the
acceleration imparted to the electron is
\begin{eqnarray}a&=&\left|{\Delta E 2G_0r\over r^3+2.504G_0\left(r+4.5G_0\Delta E\right)}\right.\nonumber
\\&-&\left.{\Delta E G_0r^2\left(3r^2+2.504G_0\right)\over \left[r^3+2.504G_0\left(r+4.5G_0\Delta E\right)\right]^2}\right|~.\end{eqnarray}
The characteristic time and length scale for the interaction when
one uses the energy $\Delta E$ for the measurement is given by
$\Delta E^{-1}$ \cite{LL}. Thus for the GUP when $\Delta E\leq
\Delta E_{cr}$ one gets
\begin{equation}\label{qclogen}\Delta x={1\over 2\Delta E}+\alpha{\left|22.536G_0^3\Delta E^5+2.504G_0^2\Delta E^3-G_0\Delta E\right|
\over \left[1+2.504G_0\left(\Delta E^2+4.5G_0\Delta
E^4\right)\right]^2}~.\end{equation}

Assuming $\alpha=1$ it is easy to check that the minimal distance
that comes from Eq.(\ref{qclogen}) $\Delta x_{min}=\Delta
x\left(\Delta E_{cr}\right)=0.147G_0^{1/2}< r_g\left(\Delta
E_{cr}\right)=4.484G_0^{1/2}$. But the object with the size
$\Delta x_{min}$ can not be black hole at all for the quantum
corrected Schwarzschild space-time does not admit the black hole
with the size less than $r_g\left(\Delta E_{cr}\right)$ \cite{BR}.
In general the parameter $\alpha$ is of order unity, but this
numerical factor can not change the result because it should be
about $1000$ the $\Delta x_{min}$ to be comparable to the
$r_g(\Delta E_{cr})$. So, one concludes that black hole
evaporation is halt due to quantum gravitational effects.  To be
strict the disappearance of the horizon beneath the mass scale
$M_{cr}$ results in the subtle question what actually is the
object left behind the evaporation, whether it is unambiguously
black hole or the classical remnant is also allowed in general.
But in the real situation this question does not arise because
when the mass approaches $M_{cr}$ the emission temperature becomes
zero \cite{BR} and due to absorption of the background radiation
this limit is simply unattainable.

Another very interesting issue that comes from effective average
action and its associated exact renormalization group equation is
the fractal structure of spacetime on sub-Planckian distances with
effective dimensionality 2 \cite{LR}. It is of interest to know if
presence of minimum uncertainty in the position considered above
($\Delta x_{min}=0.147G_0^{1/2}$) allows one to observe the
"ripples" of spacetime at sub-Planckian distances.

\vspace{0.2cm}

\centerline{\bf Acknowledgements} The author is greatly indebted
to Z.~Berezhiani, M.~Makhviladze and R.~Percacci for useful
conversations. The work was supported by the grant FEL. REG.
$980767$.


\end{document}